\documentclass{iaus}
\usepackage{graphicx}

\title[The second release of the Toru{\'n} catalogue] 
{The second release of the Toru{\'n} catalogue of Galactic post-AGB objects:
new classification scheme}

\author[R. Szczerba \etal]   
{R. Szczerba$^1$, N. Si{\'o}dmiak$^1$, G. Stasi{\'n}ska$^2$, J. Borkowski$^1$, P.
Garc{\'i}a-Lario$^3$, O. Su{\'a}rez$^{4,5}$, M. Hajduk$^1$, D.A.
Garc{\'i}a-Hern{\'a}ndez$^{6,7}$}

\affiliation{$^1$N. Copernicus Astronomical Center, Toru{\'n}, Poland \\[\affilskip]
$^2$LUTH, Observatoire de Paris, CNRS, France \\[\affilskip]
$^3$Herschel Science Centre, European Space Astronomy Centre, ESA, Spain \\[\affilskip]
$^4$Universite de Nice Sophia Antipolis, CNRS, France \\[\affilskip]
$^5$Laboratorio de Astrof{\'i}sica Espacial y F{\'i}sica Fundamental, INTA, Spain \\[\affilskip]
$^6$Instituto de Astrof{\'i}sica de Canarias, Spain \\[\affilskip]
$^7$Universidad de La Laguna, Spain}

\pubyear{2011}
\volume{283}  
\pagerange{1-2}
\jname{Planetary Nebulae: an Eye to the Future}
\editors{A.C. Editor, B.D. Editor \& C.E. Editor, eds.}
\begin{document}

\maketitle

\begin{abstract}
The investigation of post-AGB objects (proto-planetary nebulae) is very
important from the standpoint of physical and chemical changes occurring
during the late stages of stellar evolution. The {\it Toru{\'n} catalogue of
Galactic post-AGB and related objects} is an evolutive catalogue containing
astrometric, photometric and spectroscopic data as well as HST images for
all known post-AGB objects and candidates in our Galaxy. This free-access
catalogue can serve as an ideal tool to study different groups of post-AGB
objects, especially due to the fact that all information is gathered in one
place. The second release of our catalogue introduces a simple
classification scheme of post-AGB objects and includes a significant number
of new objects, photometric data, spectra and images. Here, using objects
from the catalogue
we consider the problem of the termination of the AGB phase.
\end{abstract}

\firstsection 
\section{The catalogue}
The post-AGB objects gathered in our catalogue are stars that left the AGB
but are still not hot enough to ionize the surrounding matter. This
definition does not imply that the objects will become planetary nebulae
(PN), since stellar ejecta prior to this stage may have dispersed in the
interstellar medium well before the remnant star is hot enough to ionize them.
The second release of the catalogue (http://www.ncac.torun.pl/postagb2 -
Szczerba et al. 2011, A\&A in revision) gives on-line access to about 480
likely and possible post-AGB objects, including about 110 RV Tau stars \& 70 helium
stars. There are also about 70 unlikely post-AGB objects, which sometimes are
counted in the literature as proto-PNe or post-AGBs. 

\begin{figure}[t]
\begin{center}
 \includegraphics[width=3.2in]{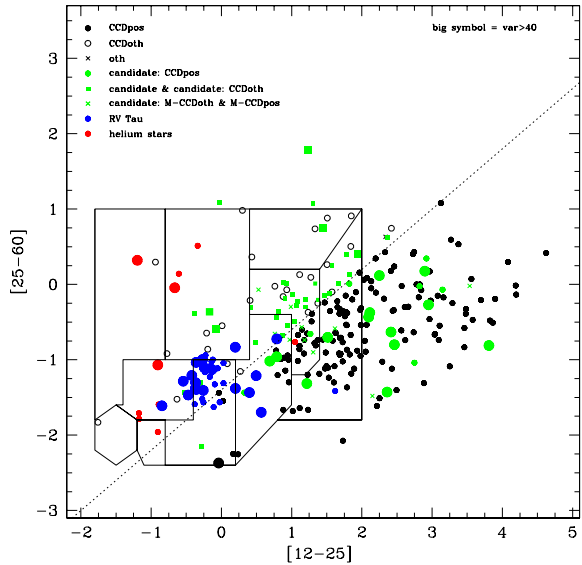} 
 \caption{IRAS color-color diagram for objects from the Toru{\'n} catalogue with IRAS fluxes.}
   \label{fig1}
\end{center}
\end{figure}

The main tool to classify a post-AGB candidate is its position in the IRAS
color-color diagram (CCD) and its IRAS variability, supplemented by some
additional criteria, if necessary. We count an object as a likely post-AGB
if it lies below the dotted line in the CCD (Fig.\,\ref{fig1}), has IRAS variability
index smaller than 40 and is not a PN (CCDpos). For objects located above
the line we used additional criteria, like: double peaked 1612 MHz OH maser
emission, luminosity class I, etc… (CCDoth). A detailed
description of the adopted criteria can be found in Szczerba et al. (2011). 
We have also distinguished a group of M-type post-AGB(?)
stars (M-CCDpos; M-CCDoth), which are discussed in Sect. 2.

The second release of the catalogue is enriched in: a) photometry from GSC
2.3, b) optical spectra from \cite[Su{\'a}rez \etal\ (2006)]{Suarez_etal06}
and \cite[Pereira \& Miranda (2007)]{PereiraMiranda07} for $\sim120$ sources, c) HST
images for $\sim100$ sources, d)
classification of spectral energy distribution (SED) according to the \cite[van der
Veen \etal\ (1989)]{Veen_etal89} scheme (additionally, we introduced class 0 for
objects with almost no IR excess), e) V \& R long-term photometric variability from
\cite[Hrivnak \etal\ (2010)]{Hrivnak_etal10} for 12 sources, f) IRAS variability
index, g) dominant chemistry –- photospheric and/or circumstellar.


\section{M-type objects and the end of AGB}
There are several objects in our catalogue with double-peaked SEDs and
spectral type M implying an effective temperature around 3000K and yet
called in the literature post-AGB objects. While double-peaked SED
suggests the end of AGB phase, the M-type contradicts this conclusion.  
However, the only M-type object (IRAS 18420-0512)
resolved by HST shows circumstellar arcs (\cite[Sahai \etal\
(2007)]{Sahai_etal07}). Therefore, we postulate that M-type objects are at
the end of their ''puffing'' phase during AGB and are just entering the
post-AGB stage of stellar evolution, so they could be called the
''transition post-AGB objects''.

\end{document}